\documentclass[submission, Proceedings]{SciPost}

\hypersetup{
    colorlinks,
    linkcolor={red!50!black},
    citecolor={blue!50!black},
    urlcolor={blue!80!black}
}

\usepackage[bitstream-charter]{mathdesign}
\DeclareSymbolFont{usualmathcal}{OMS}{cmsy}{m}{n}
\DeclareSymbolFontAlphabet{\mathcal}{usualmathcal}

\usepackage{tikz}
\usepackage{siunitx}
\usepackage{cleveref}
\usepackage{braket}

\begin{document}

\begin{center}{\Large \textbf{
Air showers and hadronic interactions with CORSIKA~8\\
}}\end{center}

\renewcommand{\thefootnote}{\alph{footnote}}

\begin{center}
Maximilian Reininghaus\textsuperscript{1} for the CORSIKA~8 Collaboration\footnote{full author list available at \url{https://tinyurl.com/corsika8-202210}}
\end{center}

\begin{center}
\textbf{1} Karlsruher Institut für Technologie, Institut für Astroteilchenphysik,\\
Postfach~3640, 76021~Karlsruhe, Germany
\\
 \url{reininghaus@kit.edu}
\end{center}

\begin{center}
\today
\end{center}


\definecolor{palegray}{gray}{0.95}
\begin{center}
\colorbox{palegray}{
  \begin{tabular}{rr}
  \begin{minipage}{0.1\textwidth}
    \includegraphics[width=23mm]{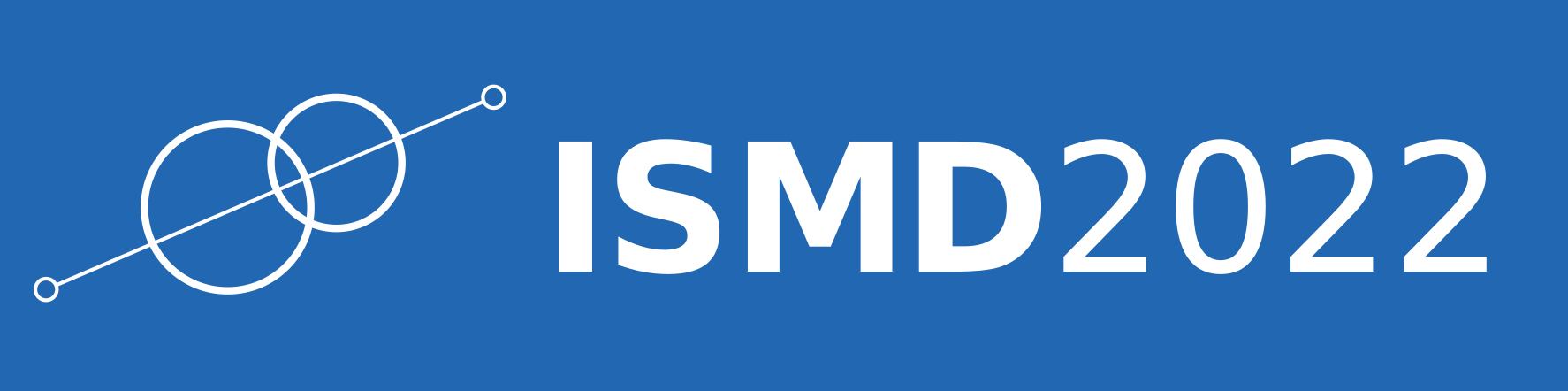}
  \end{minipage}
  &
  \begin{minipage}{0.8\textwidth}
    \begin{center}
    \textit{51st International Symposium on Multiparticle Dynamics (ISMD2022)}\\ 
    \textit{Pitlochry, Scottish Highlands, 1-5 August 2022} \\
    \doi{10.21468/SciPostPhysProc.?}\\
    \end{center}
  \end{minipage}
\end{tabular}
}
\end{center}

\section*{Abstract}
\textbf{
	The CORSIKA 8 project is a collaborative effort aiming to develop a versatile C++ framework for the simulation of extensive air showers,
	intended to eventually succeed the long-standing FORTRAN version. I present an overview of its current capabilities, focusing on aspects
	concerning the hadronic and muonic shower components. In particular, I demonstrate the “cascade lineage” feature and its application to
	quantify the importance of certain phase-space regions in hadronic interactions for muon production.
	Additionally, I show first results using Pythia 8.3, which as of late is usable as interaction model in cosmic-ray applications and is
	currently being integrated into CORSIKA 8.
}


\section{Introduction}
A large part of the astroparticle physics community deals with the measurement of extensive air showers (EAS), particle cascades
developing on macrosopically large scales of up to several \SI{10}{\km}. Making use of measurements of EAS observables in order
to infer properties of the primary particle (high-energy cosmic rays, gamma rays, neutrinos) requires accurate predictions of these.
Only Monte Caro simulations are able to provide these at the necessary level of detail, relying on numerical methods and reliable
models of the physical processes involved. The phase space covered in EAS is vast: Electromagnetic (EM), weak, and strong interactions
play a role; dozens of particle species are involved; energy scales range from $\sim\si{\kilo\eV}$ up to $\si{\zetta\eV}$.
Software tools that can keep up with these requirements are a must-have and serve as a cornerstone of the field. For more than \SI{30}{years}
the FORTRAN code CORSIKA (Cosmic Ray Simulations for KASCADE)~\cite{Capdevielle:1990,Heck:1998vt} played that role and has become a
de facto standard~\cite{Kampert:2012vi}. In recent years, however, it has become unfeasible to accomodate for the increasing needs
of upcoming experiments by continuing to extend the existing "dinosource".

Instead, efforts have been taken to develop a new C++ code from scratch, eventually termed CORSIKA~8~\cite{Reininghaus:2019jxg}, with
the goal to provide a modern, modular and flexible framework for simulations of particle showers. Since its inception in 2018~\cite{Engel:2018akg},
CORSIKA~8 is developed as open-source project by an international collaboration. The code is publicly available at the \href{https://gitlab.iap.kit.edu/AirShowerPhysics/corsika/}{gitlab repository}
and usable by early adopters. At the time of writing, a large fraction of features available in the legacy version are implemented in CORSIKA~8.
Moreover, CORSIKA~8 has a number of unique features that are not available in other codes.

In this article I focus on the capabilities in the hadronic and muonic sectors of EAS, which to date remain not fully understood, as a number
of discrepancies between experimental data and simulations suggest~\cite{Albrecht:2021yla}. More general overviews of the current state of the project can
be found in refs.~\cite{Huege:2022xbo,CORSIKA:2021czu}. The structure of this article is as follows: In \cref{sec:genealogy} I present a study of the
phase space of hadronic interactions relevant for muon production. Additionally, I show first results using Pythia~8 as hadronic interaction model
in the context of air shower simulations in \cref{sec:cr-pythia}, followed by concluding remarks.

\section{Air shower genealogy}\label{sec:genealogy}
The muon component of EAS is a tracer of the hadronic interactions happening during the shower development. 
The bulk of muons observed at ground, having energies mostly in the \SI{100}{\MeV} to \si{\GeV} range, stem from the decay of low-energy pions
and kaons that form the last generation of the hadronic cascade. Measurements of the muon content of EAS induced by ultra-high energy cosmic rays
(i.e., having energies $\geq \SI{1}{\exa\eV}$) performed in several experiments show that there is a significant excess of muons in data compared to
simulations using up-to-date versions of hadronic interaction models~\cite{Soldin:2021wyv}. It is widely believed that this discrepancy, coined
\emph{muon puzzle}~\cite{Albrecht:2021yla}, is the result of a lack of understanding and mismodelling of hadronic interactions.

\begin{figure}[bt]
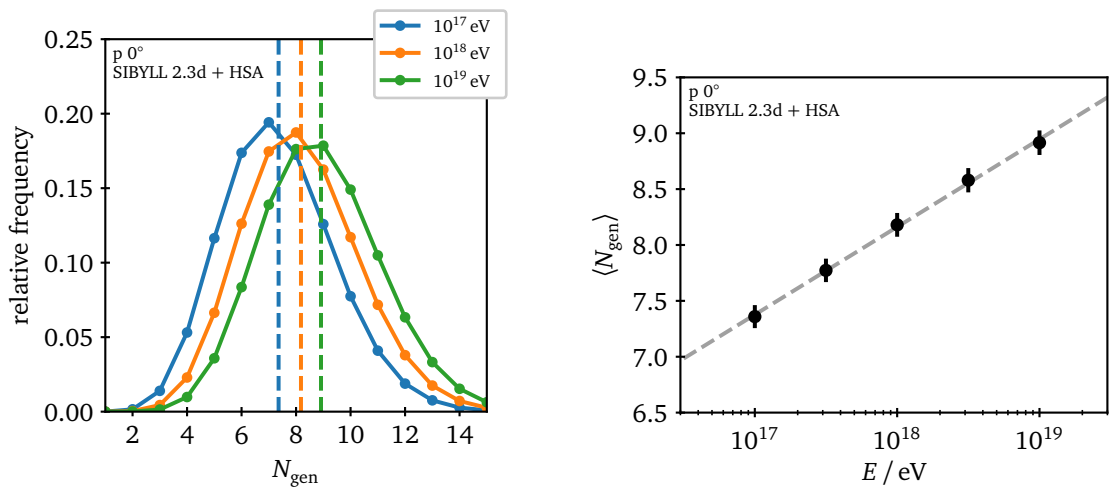

\centering
\begin{minipage}[t]{0.49\textwidth}
	\input{figures/generations-by-E0_p_0deg_sib23d+HSA.pgf}
\end{minipage}
\begin{minipage}[t]{0.49\textwidth}
	\input{figures/Ngen_vs_E_sib23d+HSA_p_0deg.pgf}
\end{minipage}
\caption{Left: Distributions of the number of muon ancestor generations for primary energies of $E_0 = 10^{17}, 10^{18}, 10^{19}\,\si{\eV}$.
Vertical dashed lines indicate the mean value of the distribution of the same colour. Taken from ref.~\cite{Reininghaus:2021zge}.
Right: Mean number of muon ancestor generations as function of primary energy. Taken from ref.~\cite{Reininghaus:2021}.}
\label{fig:Ngen}
\end{figure}

CORSIKA~8 is particularly well suited to shed more light onto the muon puzzle: It allows keeping the complete \emph{lineage} of particles in memory
so that particles can be related to any of their predecessor generations up to the primary particle. Details on the technical implementation are given
in refs.~\cite{Alves:2021wiw,Reininghaus:2021}. The first study to exploit this information is presented in ref.~\cite{Reininghaus:2021zge}, whose
results I summarize here. \Cref{fig:Ngen} shows results regarding the number of generations $N_{\mathrm{gen}}$, i.e.\ the number of hadronic interactions
happening between the primary particle and the final muon that reaches ground. It is an important quantity because the number of muons grows exponentially
with $N_{\mathrm{gen}}$ and small errors in the modelling of these interactions get amplified $N_{\mathrm{gen}}$ times, leading to a potentially large
impact on the muon number~\cite{Cazon:2020jla}. The left plot shows the $N_{\mathrm{gen}}$ distributions of proton-induced EAS with energies of
$E_0 = 10^{17}, 10^{18}, 10^{19}\,\si{\eV}$, simulated using the interaction model SIBYLL~2.3d~\cite{Riehn:2019jet} at high energies ($> \SI{63.1}{\GeV}$)
together with the Hillas Splitting algorithm (HSA)~\cite{Hillas:1981} for low energies. With increasing primary energy, the distributions shift
towards higher values of $N_{\mathrm{gen}}$. The right plot shows the dependence of the mean $\braket{N_{\mathrm{gen}}}$ on the primary energy.
The logarithmic behaviour follows what is expected from the Heitler--Matthews toy model~\cite{Matthews:2005sd}. A more detailed analysis is given in
ref.~\cite{Reininghaus:2021}.

\begin{figure}[t]
\centering
\input{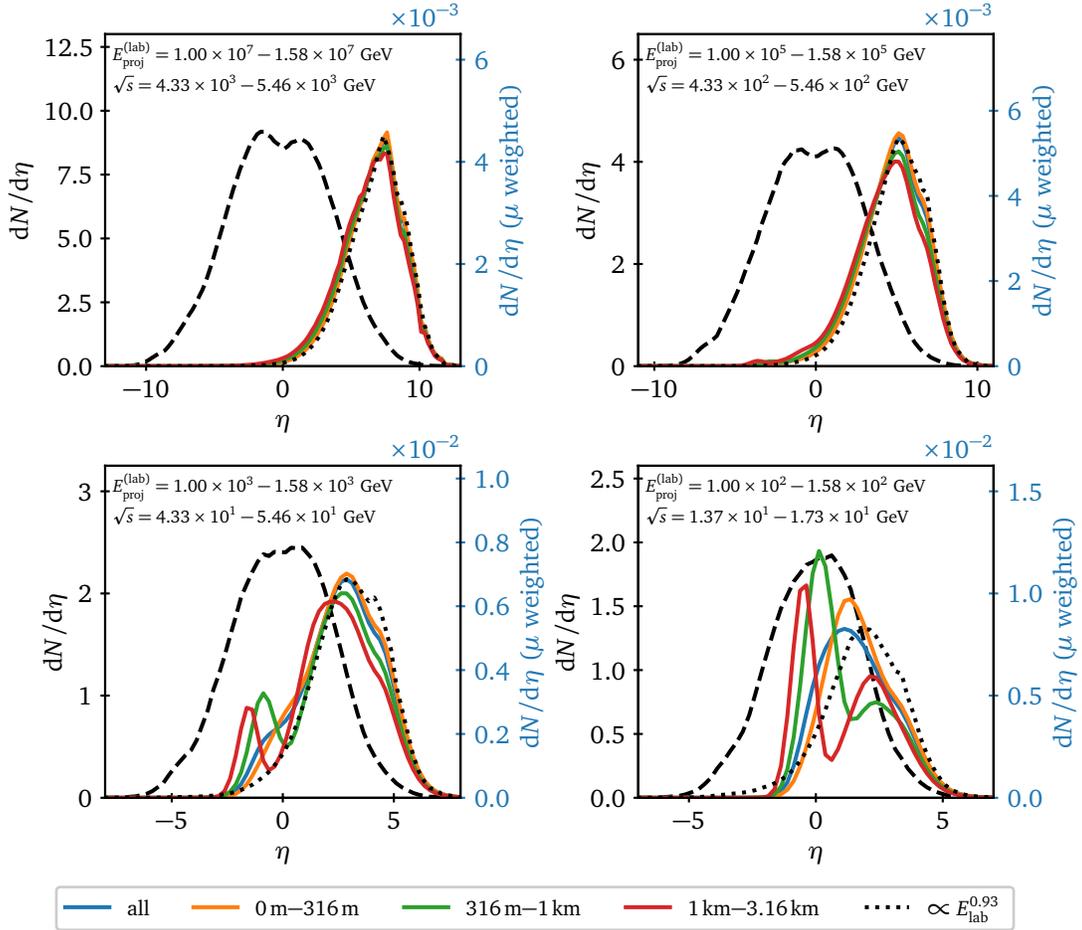}
\caption{Pseudorapidity distributions of $\pi^\pm + \text{air} \rightarrow \text{charged hadrons}$.
The dashed line indicates generator-level distributions while the coloured and dotted lines show muon-weighted distributions of
vertical, proton-induced showers of $10^{19}\,\si{\eV}$ (in arbitrary units). Taken from ref.~\cite{Reininghaus:2021zge}.}
\label{fig:eta}
\end{figure}

A second study deals with the importance of different phase-space regions in hadronic interactions with respect to muon production.
\Cref{fig:eta} shows pseudorapidity ($\eta$) distributions of charged hadrons in $\pi^\pm$-air collisions in four energy bins. Besides the pure generator-level
spectra (dashed lines), which are in principle measurable in accelerator experiments, the \emph{muon weighted} distributions are superimposed (coloured lines).
This muon weight is given by the number of muons stemming directly or indirectly from the secondaries emitted at a given $\eta$, potentially after applying
some selection criterion. The plot shows that at sufficiently high interaction energies ($\sqrt s \gtrsim \SI{100}{\GeV}$) essentially only particle production in the forward region plays a role,
irrespective of the lateral distance of the muons. In this regime the muon weight can also be estimated well from the Heitler--Matthews model~\cite{Albrecht:2021yla}, indicated
by the dotted line. At low interaction energies ($\sqrt s \lesssim \SI{50}{\GeV}$), however, the central region gains relevance especially for muons at distances larger than
a few \SI{100}{\metre}.

\section{Towards Pythia~8 as interaction model in EAS simulations}\label{sec:cr-pythia}
A currently ongoing development is the inclusion of Pythia~8.3~\cite{Bierlich:2022pfr} into CORSIKA~8 as hadronic interaction model.
Until recently, this was technically unfeasible, but since the latest release (Pythia~8.307) a number of new features allow for an easy
use in EAS simulations~\cite{Sjostrand:2021dal}: a) the ability to generate single events at arbitrary energies without a time-consuming re-initialization
of the model, b) a much wider range of possible projectile species, c) an extended range of interaction energies down to \SI{200}{\MeV} (lab-frame),
d) a simplified model of nuclear matter, allowing hadron-air collisions.

\begin{figure}[t]
\centering
\input{figures/xmax-nmu_py8.pgf}
\caption{Mean shower maximum $\braket{X_{\mathrm{max}}}$ vs.\ number of muons $N_{\mu}$ for showers at $10^{17.5}\,\si{\eV}$ and \ang{60} inclination}
\label{fig:XmaxNmu}
\end{figure}

For a first comparison we have simulated showers with an energy of $10^{17.5}\,\si{\eV}$ and an inclination
of \ang{60}. Hadrons and muons are fully propagated with CORSIKA~8, while EM particles are redirected into the CONEX code~\cite{Bergmann:2006yz}, which
simulates the EM component of the shower using a numerical solution of the cascade equations describing the longitudinal development. We use
QGSJet-II.04~\cite{Ostapchenko:2010vb}, SIBYLL~2.3d and Pythia~8.307 as high-energy interaction models. In each case Pythia is used as low-energy
interaction model. \Cref{fig:XmaxNmu} shows the results regarding the mean shower maximum $\braket{X_{\mathrm{max}}}$ and $N_{\mu}$.
While $N_{\mu}$ obtained with Pythia is in the same ballpark as the other models, a deviation of $\braket{X_{\mathrm{max}}}$ as large as the difference
between proton and helium is apparent. Further studies to explain these differences, stemming from differences in hadron-air cross-sections, are
ongoing and will be presented elsewhere~\cite{Reininghaus:2022}. Note that in our setup Pythia~8.307 cannot be used for nucleus-nucleus collisions
at the moment so that we consider only proton-induced showers in that case.

\section{Conclusion}
CORSIKA~8 is a modern framework for the simulation of particle showers in air and other media. Over the past few years, a lot of progress has been
made to make it a reliable, versatile and future-proof tool. Although not yet feature-complete, in some aspects it can be used for studies that
have previously been impossible. The availability of the complete particle lineage allows detailed studies of hadronic interactions and their
relevance for muon production, which helps to shed light on the muon puzzle. It emphasizes the importance of dedicated accelerator measurements
at both high and low energies that cover the relevant phase-space.

The ongoing integration of Pythia~8 as new hadronic interaction model in CORSIKA~8 offers a number of new perspectives for EAS simulations.
On the one hand, the reduced energy threshold of only \SI{200}{\MeV} renders it suitable for both low and high energies. On the other hand,
Pythia~8 can be tuned by the users, allowing to study the impact of internal model parameters on EAS observables, or possibly even to conduct
combined fits to accelerator measurements and EAS data at the same time.

\section*{Acknowledgements}
I thank Torbjörn Sjöstrand and Marius Utheim for providing example code that greatly simplified the integration of
Pythia~8 in CORSIKA~8.

Simulations were performed on the bwForCluster BinAC of the University of Tübingen, 
supported by the state of Baden-Württemberg through bwHPC and the DFG through grant
no.\ INST~37/935-1~FUGG.

\bibliography{literature.bib}

\nolinenumbers

\end{document}